\documentclass[prl,twocolumn,aps,showpacs]{revtex4}
\usepackage{graphicx,color}
\usepackage{epstopdf}
\usepackage{latexsym,amssymb}

\def\be{\begin{equation}}
\def\ee{\end{equation}}

\def\bc{\begin{center}}
\def\ec{\end{center}}

\begin{document}

\title{Origin of folded bands in metamaterial crystals}

\author{Peter Marko\v{s} and Richard Hlubina} 

\affiliation{Department of Experimental Physics, Comenius University,
  Mlynsk\'{a} Dolina F2, 842 48 Bratislava, Slovakia}

\begin{abstract}
Recently it has been found numerically that the spectra of
metamaterial crystals may contain pairs of bands which disappear
inside the Brillouin zone. We observe that the wave equations for such
systems are essentially non-Hermitian, but ${\cal P}{\cal
  T}$-symmetric. We show that the real-frequency spectra correspond to
${\cal P}{\cal T}$-symmetric solutions of the wave quation. At those
momenta in the Brillouin zone where apparently no solutions exist,
there appear pairs of complex-frequency solutions with spontaneously
broken ${\cal P}{\cal T}$ symmetry.
\end{abstract}


\maketitle

One of the basic characteristics of waves propagating in material
media is their frequency spectrum. In periodic systems, for instance
in photonic crystals, the frequency $\omega$ is a function of the wave
vector $\vec{q}$, $\omega=\omega_s(\vec{q})$, where $\vec{q}$ is
restricted to an elementary tile of the reciprocal space, the
so-called first Brillouin zone.  The discrete index $s$ numerates
distinct branches of the dispersion, which correspond to different
distributions of the wave field within the unit cell of the periodic
system. Since in macroscopic systems the wave vector $\vec{q}$ changes
quasi-continuously, each branch $s$ leads in general to a finite
interval of allowed frequencies, the so-called bands, which may be
divided by band gaps in between them \cite{sakoda}.

In most systems studied so far, either in the solid-state or photonic
context, for each of the branches the function
$\omega=\omega_s(\vec{q})$ stretches throughout the whole Brillouin
zone. This is a simple consequence of Hermiticity. In fact, the plane
wave $e^{i\vec{q}\cdot\vec{x}}$ may experience a Bragg scattering to
any of the plane waves of the form
$e^{i(\vec{q}+\vec{K})\cdot\vec{x}}$, where $\vec{K}$ is a reciprocal
lattice vector. In the basis of such states, the Schr\"odinger or wave
equation takes the form of an eigenvalue problem
$H_{\vec{K}\vec{K^\prime}}(\vec{q})c_{\vec{K^\prime}}=\lambda(\vec{q})
c_{\vec{K}}$. For a fixed cut-off we are then dealing with an $N\times
N$ matrix which, if it is Hermitian, is guaranteed to have $N$ real
eigenvalues, independently of the value of $\vec{q}$. Smooth changes
of $H_{\vec{K}\vec{K^\prime}}(\vec{q})$ lead then to smooth changes of
$\lambda(\vec{q})$, resulting in bands which cannot disappear inside
the Brillouin zone. In other words, the number of eigenfrequencies
cannot be reduced in a certain interval of wave vectors.

\begin{figure}[b]
\begin{center}
\includegraphics[width=0.55\linewidth]{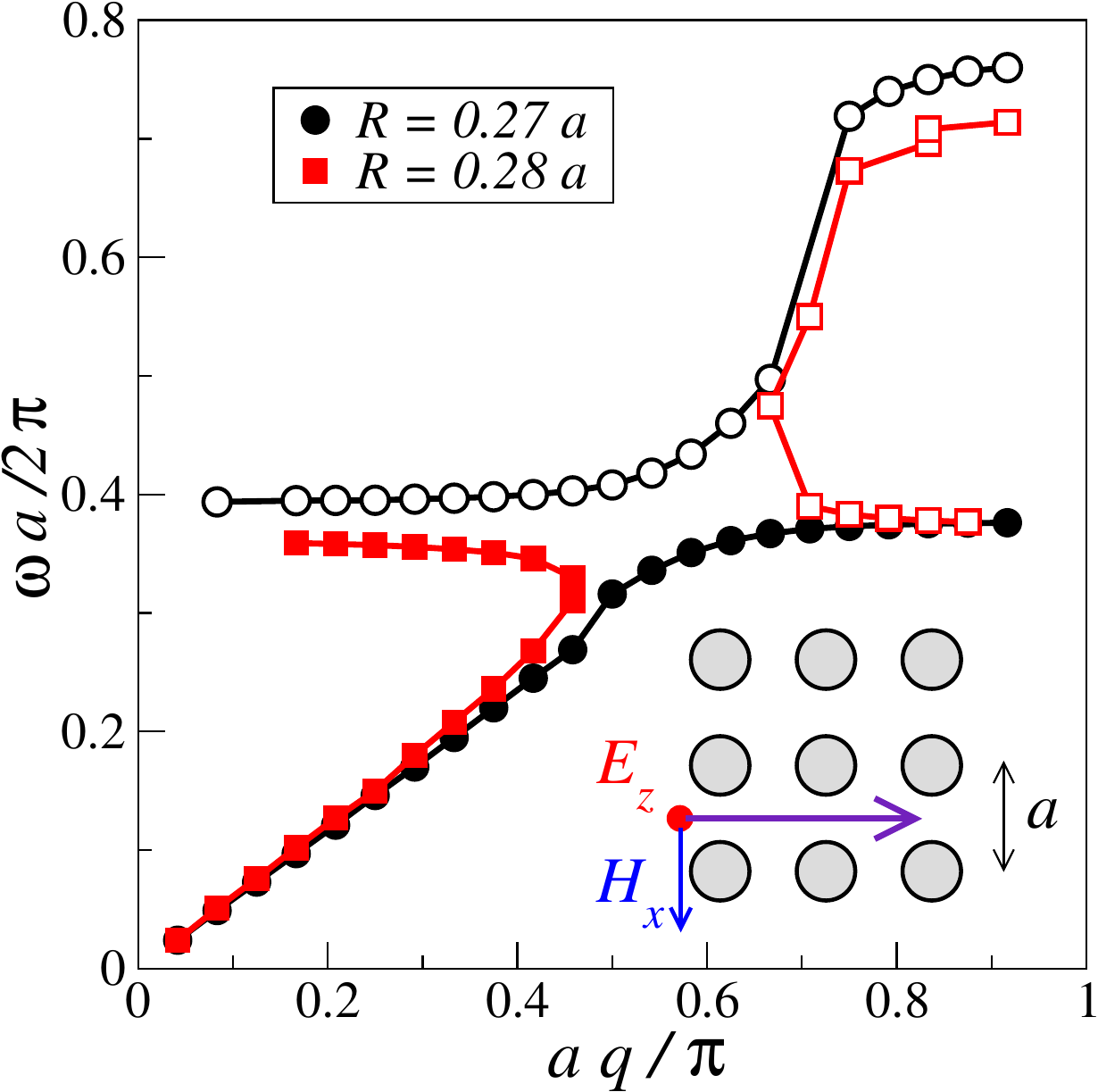}
\end{center}
\caption{(Color online) Dispersion relation $\omega=\omega(\vec{q})$
  in the $\Gamma X$ direction for a two-dimensional photonic crystal
  made of metamaterial cylinders with real and dispersionless
  permittivity $\varepsilon=-1.8$ and permeability $\mu=-5$, see
  inset. Electric field is taken to be parallel to the cylinders. The
  two sets of curves correspond to cylinder radii slightly above and
  below the critical radius $R_c\approx 0.275a$ \cite{pm}. }
\label{fig:intro}
\end{figure}

However, in numerical simulations it has recently been found that in
certain systems the bands may disappear inside the Brillouin zone,
forming the so-called folded bands \cite{chen}. In particular, such
behavior has been observed in photonic crystals in the form of a
square array of metamaterial cylinders immersed in vacuum. As an
example, in Fig.~\ref{fig:intro} we show the two lowest-frequency bands for
such a metamaterial photonic crystal calculated numerically from
transmission spectra \cite{pm}.  Folded bands appear when the radius
of cylinders $R$ increases above the critical value $R_c\approx
0.275a$, where $a$ is the spatial period of the crystal.  This
surprising result indicates that the wave equation for electromagnetic
field in a metamaterial photonic crystal is non-Hermitian \cite{john}.

Actually, it is well known that if both, permittivity $\varepsilon$
and permeability $\mu$, are non-constant functions of the spatial
coordinate, then the wave equation for the magnetic field $\vec{H}$
reads
\begin{equation}
{\cal M}\vec{H}\equiv\mu^{-1}
\textrm{rot}\left[\varepsilon^{-1}\textrm{rot}\vec{H}\right]
= \omega^2\vec{H},
\label{eq:wave_eq}
\end{equation}
where the operator ${\cal M}$ is non-Hermitian even in ordinary
photonic crystals made from dissipationless components
\cite{joan-pc,sakoda}. Note that we set the speed of light $c=1$.

So how can it be that folded bands have not been observed in ordinary
photonic crystals? The reason is that the non-Hermitian character of
Eq.~(\ref{eq:wave_eq}) is often not essential, since it can be avoided
by a reformulation of the problem. For instance, if the permeability
$\mu$ is real and positive definite, one can redefine the magnetic
field by $\vec{H}=\vec{h}/\sqrt{\mu}$, thereby transforming
Eq.~(\ref{eq:wave_eq}) to the form
\begin{equation}
{\cal O}\vec{h}\equiv \mu^{-1/2}
\textrm{rot}\left[\varepsilon^{-1}\textrm{rot}
\left(\mu^{-1/2}\vec{h}\right)\right]
= \omega^2\vec{h}
\label{eq:wave_eq2}
\end{equation}
with an explicitly Hermitian operator ${\cal O}$
\cite{sakoda}. 

The observation of folded bands in metamaterial photonic crystals
therefore suggests that their non-Hermitian character should be {\it
  essential}, i.e. not avoidable by any reformulation. In particular,
it will be shown later that, in presence of interfaces between
ordinary dielectric regions where $\sqrt{\mu}$ is purely real and
metamaterial regions with purely imaginary $\sqrt{\mu}$, the operator
${\cal O}$ remains non-Hermitian \cite{note}.

The goal of the present paper is to demonstrate that the appearance of
folded bands in metamaterial photonic crystals is a direct consequence
of their essential non-Hermiticity. To this end, we will start by
studying the simplest possible crystal structure, namely a
one-dimensional (1D) periodic stack of right- and left-handed
materials. Several anomalous features of electromagnetic wave
propagation have already been observed in this model
\cite{Nefedov02,Li03,Wu03,Bria04}. Here we observe that the 1D model
exhibits the so-called ${\cal P}{\cal T}$ symmetry \cite{Bender05}
and, making use of this recently developed concept, we will explain
the presence of folded bands in the spectrum of this model.  Similar
reasoning will be later applied to two-dimensional (2D) metamaterial
crystals studied in \cite{chen,pm}.

\begin{figure}[t]
\begin{center}
\includegraphics[width=0.75\linewidth]{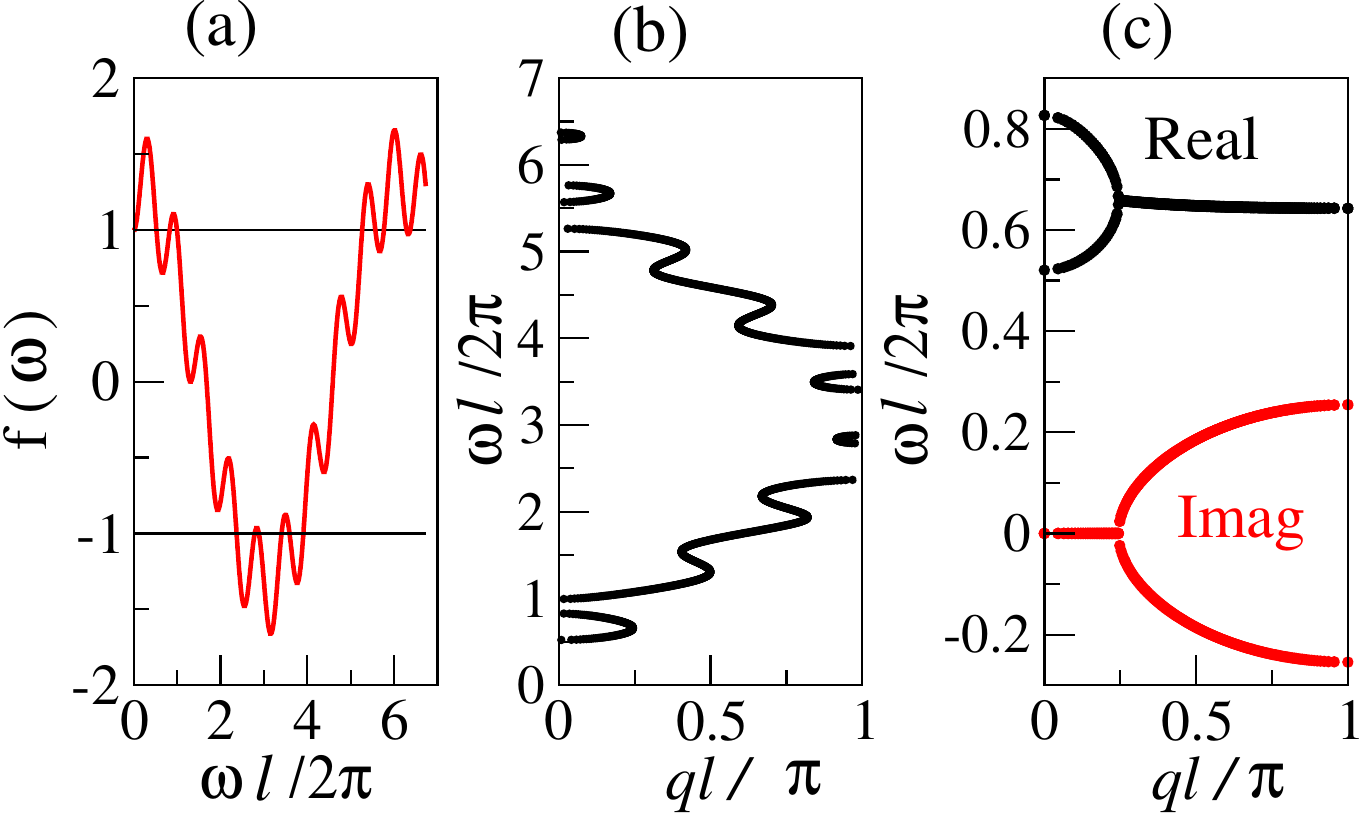}
\end{center}
\caption{(Color online) 1D model with $\varepsilon_a=1$, $\mu_a=1$,
  $\ell_a=1$, and $\varepsilon_b=-9$, $\mu_b=-1$, $\ell_b=0.41$.  (a)
  Function $f(\omega)$. (b) Frequency spectrum. (c) Real and imaginary
  parts of frequency in the lowest folded band.}
\label{fig:1d}
\end{figure}

We assume that the 1D stack consists of materials $a$ and $b$
characterized by $\varepsilon_i$, $\mu_i$, refractive indices
$n_i=\sqrt{\varepsilon_i\mu_i}$, impedances
$Z_i=\sqrt{\mu_i/\varepsilon_i}$, and thicknesses $\ell_i$, where
$i=a,b$.  All material parameters are assumed to be real and
frequency-independent.  The frequency spectrum of transverse
electromagnetic waves, which propagate perpendicularly to the slabs
with wave vector $q$, can be determined from the implicit equation
\cite{Yeh05}
\begin{equation}
f(\omega)\equiv
\frac{1}{2}(A+1)\cos\omega\tau_{+}
-\frac{1}{2}(A-1)\cos\omega\tau_{-}
=\cos q\ell,
\label{eq:1d}
\end{equation}
where $\ell=\ell_a+\ell_b$ is the length of the unit cell,
$\tau_{\pm}=\tau_a \pm\tau_b$ with $\tau_i=n_i\ell_i$, and $A
=(Z_a/Z_b+Z_b/Z_a)/2 >1$ is the impedance mismatch between the slabs
$a$ and $b$.

In ordinary photonic crystals with $\varepsilon_i>0$ and $\mu_i>0$ we
have $\tau_{+}>|\tau_{-}|\geq 0$. Therefore the larger-amplitude first
term of $f(\omega)$ oscillates faster than the smaller-amplitude
second term. Let us denote the positions of local extrema of the
function $f(\omega)$ as $\omega^\ast$. One can check easily
\cite{proof} that $|f(\omega^\ast)|\geq 1$ and from here it follows
that no folded bands can be present in the spectrum. This was of
course to be expected, since the wave equation of an ordinary photonic
crystal can be Hermitized.

Now let us assume that the slab $a$ is an ordinary dielectric with
$\varepsilon_a>0$ and $\mu_a>0$, whereas the slab $b$ is made from a
metamaterial with $\varepsilon_b<0$, $\mu_b<0$, and $n_b<0$. In this
case $\tau_{-}>|\tau_{+}|\geq 0$ and it is the smaller-amplitude
second term of $f(\omega)$ which oscillates faster than the
larger-amplitude first term. As shown explicitly in
Fig.~\ref{fig:1d}(a), then the values of $f(\omega^\ast)$ may lie
within the interval $(-1,1)$, and as a result folded bands can form in
the spectrum. Such folded bands have been observed previously for
oblique wave propagation \cite{Bria04,zeron}.  Similarly as in the
case of 2D metamaterial photonic crystals \cite{pm}, also in the 1D
case folded bands form only in a subset of the parameter space $(A,
\beta = \tau_{-}/\tau_{+})$ of Eq.~(\ref{eq:1d}).

Since Eq.~(\ref{eq:1d}) does not shed light on the mathematical
structure of the 1D problem, we shall restate its basic equations. The
wave equation Eq.~(\ref{eq:wave_eq}) reads piecewise
\begin{equation}
-n_i^{-2}H^{\prime\prime}=\omega^2 H.
\label{eq:problem1}
\end{equation} 
If we take the center of the slab $a$ as the origin, then the boundary
conditions at the interfaces $\xi=\pm\ell_a/2$ between the slabs
require
\begin{equation}
H(\xi_{-})=H(\xi_{+}),
\qquad
E(\xi_{-})=E(\xi_{+}),
\label{eq:problem2}
\end{equation}
where $\xi_{-}$ and $\xi_{+}$ are infinitesimally shifted from $\xi$
to the left and right, respectively, and
$E(\xi_{\pm})=H^\prime(\xi_{\pm})/\varepsilon(\xi_{\pm})$.  Moreover,
from the Bloch theorem follows an additional boundary condition
\begin{equation}
H(\ell/2)=e^{iq\ell}H(-\ell/2).
\label{eq:problem3}
\end{equation}
The boundary-value problem which we have to solve is defined by
Eqs.~(\ref{eq:problem1},\ref{eq:problem2},\ref{eq:problem3}). 

Let us prove that the operator ${\cal O}$ for the 1D problem defined
by Eqs.~(\ref{eq:problem1},\ref{eq:problem2},\ref{eq:problem3}) is not
Hermitian. To this end, let us define magnetic and ``electric'' fields
corresponding to $h(x)$, $H_h(x)=h(x)/\sqrt{\mu(x)}$ and
$E_h(x)=H_h^\prime(x)/\varepsilon(x)$.  If ${\cal O}$ was Hermitian,
then the quantity $D=\int_{-\ell/2}^{\ell/2}dx \left[ g^\ast(x){\cal
    O}h(x)- \left({\cal O}g(x)\right)^\ast h(x)\right]$ should be zero
for any pair of functions $h(x)$ and $g(x)$, both of which generate
fields $H_h$, $E_h$, and $H_g$, $E_g$ satisfying the boundary
conditions Eqs.~(\ref{eq:problem2},\ref{eq:problem3}).  Integrating
per parts and taking proper care of the boundary terms, we find that
$D=\Delta(-\ell_a/2)+\Delta(\ell_a/2)$, where
\begin{eqnarray*}
\Delta(\xi)&=&
H_h(\xi)\left[E_{g^\ast}(\xi_{-})-E_{g^\ast}(\xi_{+})\right]
\\
&+&E_h(\xi)\left[H_{g^\ast}(\xi_{+})-H_{g^\ast}(\xi_{-})\right]
\end{eqnarray*}
are the interface contributions.  Note that in a right-handed medium
$[H_g(x)]^\ast=H_{g^\ast}(x)$ and $[E_g(x)]^\ast=E_{g^\ast}(x)$,
whereas in a left-handed medium $[H_g(x)]^\ast=-H_{g^\ast}(x)$ and
$[E_g(x)]^\ast=-E_{g^\ast}(x)$.  It follows that in an ordinary
photonic crystal the boundary conditions Eq.~(\ref{eq:problem2}) imply
that $E_{g^\ast}(\xi_{-})=E_{g^\ast}(\xi_{+})$ and
$H_{g^\ast}(\xi_{+})=H_{g^\ast}(\xi_{-})$. Therefore $D=0$, as was to
be expected. However, in a metamaterial photonic crystal
$E_{g^\ast}(\xi_{-})=-E_{g^\ast}(\xi_{+})$ and
$H_{g^\ast}(\xi_{+})=-H_{g^\ast}(\xi_{-})$. Thus $D\neq 0$ and hence
the operator ${\cal O}$ is non-Hermitian.

\begin{figure}[t]
\begin{center}
\includegraphics[width=0.55\linewidth,angle=90]{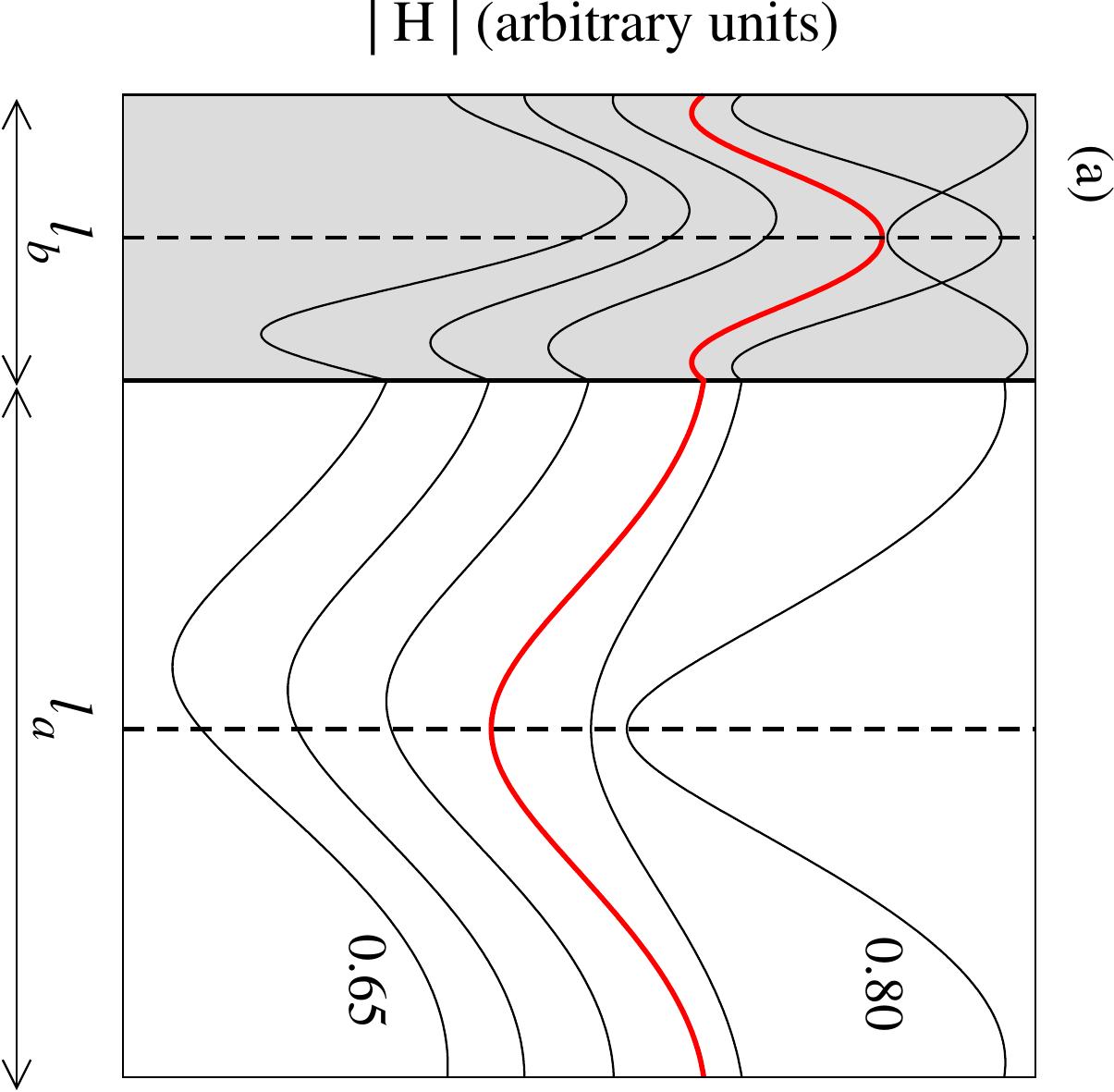}
~~
\includegraphics[width=0.21\linewidth,angle=0]{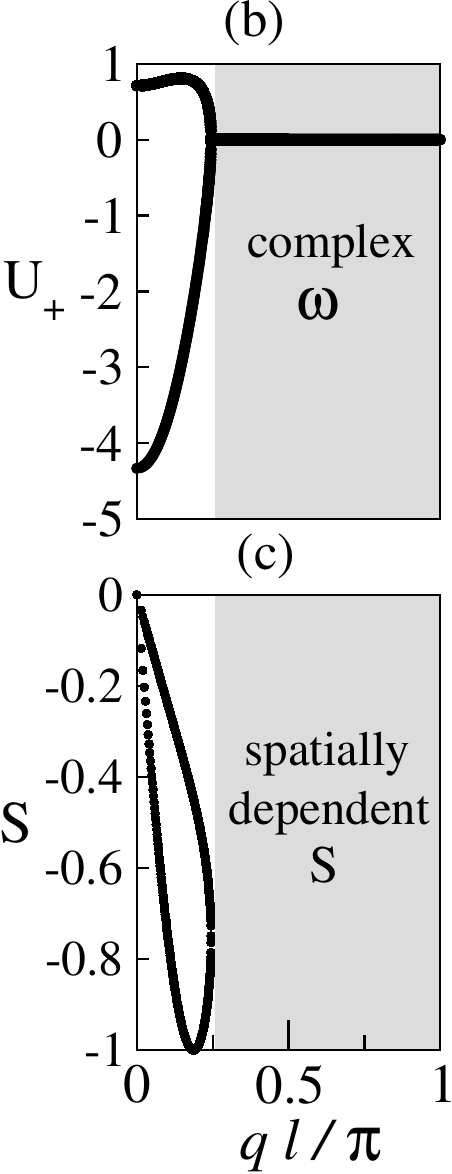}
\end{center}
\caption{(Color online) Observables for the 1D model with the same
  parameters as in Fig.~\ref{fig:1d}. (a) Spatial distribution of
  $|H(x)|$. White and shaded regions correspond to layers $a$ and $b$,
  respectively. The vertical dashed lines denote the centers of the
  layers.  Field distributions for the lowest folded band are shown
  for various values of $\cos q\ell$, from bottom to top: 0.650,
  0.700, 0.710, 0.719 (critical value), 0.730 and 0.800.  Data are
  verticaly shifted for clarity. In ${\cal P}{\cal T}$-symmetric
  solutions the field $|H(x)|$ is even with respect to the center of
  any layer. ${\cal P}{\cal T}$ symmetry-breaking solutions lack this
  symmetry \cite{note2}. (b) The integral quantity $U_{+}$ as a
  function of $q$. Note that $U_{+}=0$ for ${\cal P}{\cal T}$
  symmetry-breaking solutions. (c) Real part of the Poynting vector,
  $S$, as a function of $q$. In all panels we use the normalization
  $|H(\ell_a/2)|=1$.}
\label{fig:field}
\end{figure}

Therefore the next question we should ask is: if the boundary value
problem Eqs.~(\ref{eq:problem1},\ref{eq:problem2},\ref{eq:problem3})
is non-Hermitian, how can its spectrum be real at all?

In order to answer this question, let us define operators of
``parity'' ${\cal P}$ and ``time-reversal'' ${\cal T}$ by $({\cal
  P}H)(x)=H(-x)$ and $({\cal T}H)(x)=H^\ast(x)$, respectively.  We
find that Eqs.~(\ref{eq:problem1},\ref{eq:problem2}) are invariant
under the action of both ${\cal P}$ and ${\cal T}$, while
Eq.~(\ref{eq:problem3}) is invariant only under the combined
antilinear operator ${\cal P}{\cal T}$. Therefore the 1D model is
${\cal P}{\cal T}$-symmetric \cite{optics}, as also observed for a
somewhat similar boundary-value problem studied in \cite{Krejcirik06}.

Since ${\cal P}^2=1$ and ${\cal T}^2=1$, the eigenvalues of the
antilinear operator ${\cal P}{\cal T}$ are $\lambda=e^{i\varphi}$,
where $\varphi$ is real \cite{Bender05}.  Therefore a ${\cal P}{\cal
  T}$-symmetric solution should satisfy 
\begin{equation}
H^\ast(-x)=e^{i\varphi} H(x).
\label{eq:symmetry}
\end{equation}
If the ${\cal P}{\cal T}$ symmetry is not broken, i.e. if the
eigenstates of the wave equation are simultaneously also eigenstates
of ${\cal P}{\cal T}$, the eigenvalues $\omega^2$ have to be real
\cite{Bender05}.  On the other hand, if the eigenstates of the wave
equation spontaneously break the ${\cal P}{\cal T}$ symmetry, then the
eigenfrequencies corresponding to $H_1(x)$ and $H_2(x)={\cal P}{\cal
  T}H(x)$ become complex conjugate. This is explicitly shown in
Fig.~\ref{fig:1d}(c), which has been obtained by solving
Eq.~(\ref{eq:1d}) under the assumption of a complex frequency
$\omega=\omega_1+i\omega_2$.  Such pairs of eigenvalues have been
observed previously in the 1D problem \cite{Wu03} and also in other
${\cal P}{\cal T}$-symmetric problems \cite{Mostafazadeh11}. Note that
the complex-frequency solutions appear exactly for those values of
momentum $q$, where the band folding has been observed. Therefore, if
we allow for complex-valued frequencies, the number of eigenvalues
does not vary with $q$.

We have checked numerically that for real-frequency solutions, the
magnetic field does obey Eq.~(\ref{eq:symmetry}). On the other hand,
according to Eq.~(\ref{eq:symmetry}), a ${\cal P}{\cal T}$-symmetric
solution should satisfy $|H(-x)|=|H(x)|$.  But Fig.~\ref{fig:field}(a)
clearly shows that this criterion is not satisfied for
complex-frequency solutions, thereby proving explicitly that they
break the ${\cal P}{\cal T}$ symmetry.

\begin{figure}[t]
\begin{center}
\includegraphics[width=0.8\linewidth]{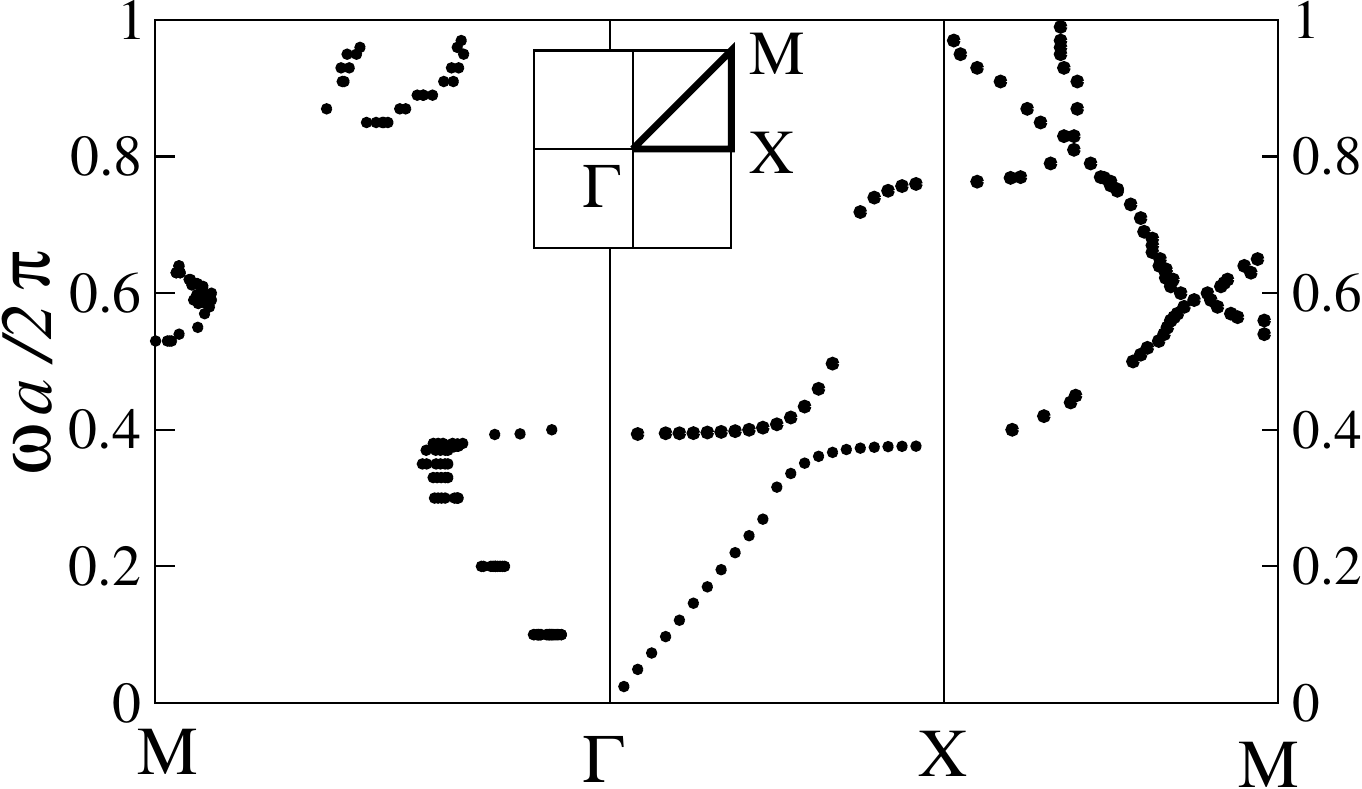}
\end{center}
\caption{Dispersion relations for the 2D model with cylinder radius
  $R=0.27a$ along special lines in the 2D Brillouin zone (see
  inset). Electric field is polarized along the cylinders. Note that
  in the $\Gamma M$ direction, the two lowest bands are already
  folded. In the $X M$ and $\Gamma M$ directions, a third band enters
  the studied frequency range.}
\label{fig:2d}
\end{figure}

Finally, we will turn back to the case of the square array of
metamaterial cylinders with lattice constant $a$.  Let us study
electromagnetic waves with the electric field parallel with the
cylinders and the $z$ axis.  Then the wave equation reads piecewise
$n_i^{-2}\triangle E=\omega^2 E$ with a 2D Laplacian $\triangle$. The
boundary conditions require the continuity of $E(x,y)$ and
$E^\prime_n(x,y)/\mu(x,y)$ across the cylinder surfaces, where
$E^\prime_n$ denotes a normal derivative. Moreover, a Bloch wave with
wave vector $\vec{q}=(q_x,q_y)$ should satisfy the boundary conditions
\begin{eqnarray}
E(a/2,y)&=&e^{iq_xa}E(-a/2,y),
\nonumber
\\
E(x,a/2)&=&e^{iq_ya}E(x,-a/2).
\label{eq:Bloch}
\end{eqnarray}

Let us define the following 2D generalizations of the operators of
``parity'' ${\cal P}$ and ``time-reversal'' ${\cal T}$: $({\cal
  P}E)(x,y)=E(-x,-y)$ $({\cal T}E)(x,y)=E^\ast(x,y)$.  Since the
geometrical structure is ${\cal P}$-invariant and since the material
parameters $\varepsilon_i,\mu_i$ and $n_i$ are taken to be real, the
wave equation and the boundary conditions inside the unit cell are
invariant under both, parity ${\cal P}$ and time-reversal ${\cal T}$.
However, similarly as in the 1D example studied before, the Bloch
boundary condition Eq.~(\ref{eq:Bloch}) is preserved only under the
action of the combined operator ${\cal P}{\cal T}$. Consequently, our
2D boundary value problem is again invariant under the action of the
operator ${\cal P}{\cal T}$.

Once the non-Hermiticity and ${\cal P}{\cal T}$ symmetry of the 2D
problem have been established, the reality of the spectrum in the
${\cal P}{\cal T}$-symmetric sector of eigenstates follows. In analogy
with the 1D case, we conjecture that in those regions of the Brillouin
zone where the folded bands disappear, there exist pairs of
complex-frequency solutions in the spectrum. Unfortunately, since such
states can not be seen in a numerical transmission experiment
\cite{pm}, we can not present a direct proof of their existence.

\begin{figure}[t]
\begin{center}
\includegraphics[width=0.7\linewidth]{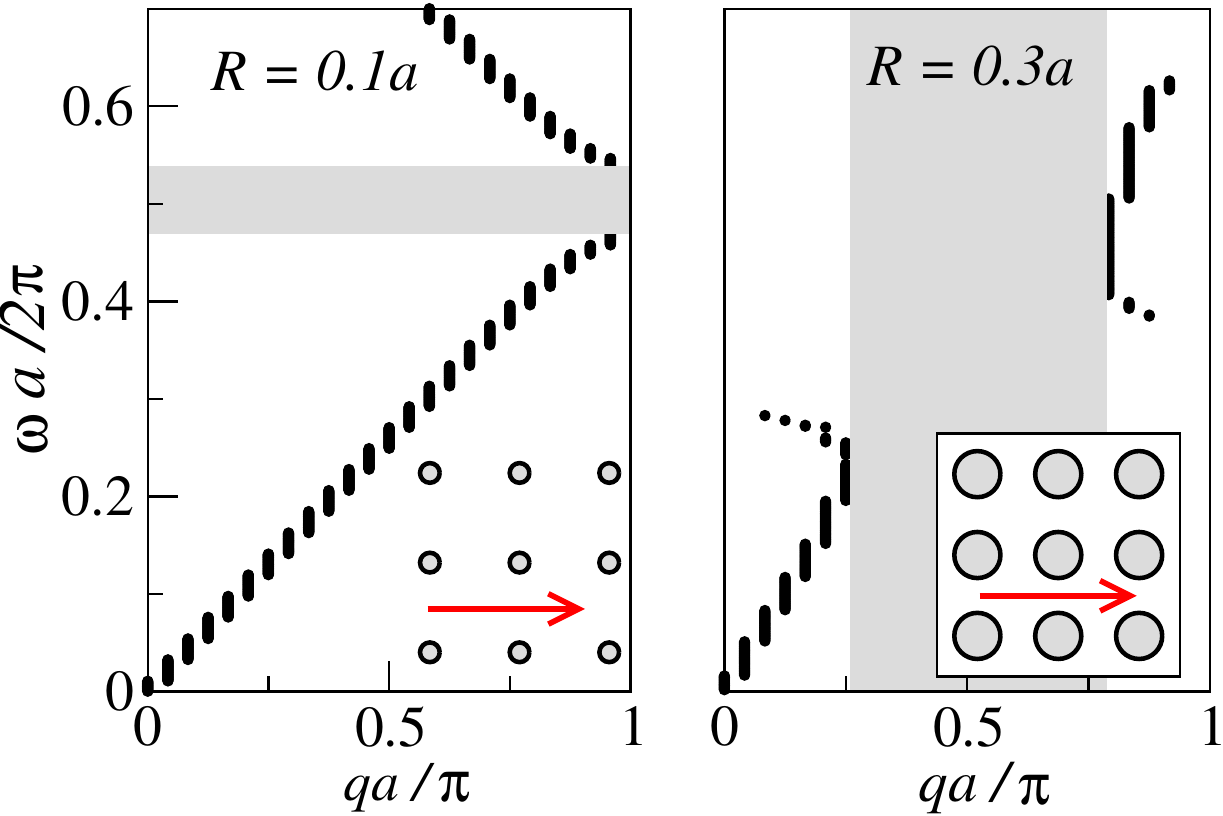}
\end{center}
\caption{(Color online) Dispersion relations along $\Gamma X$ for the
  2D model with cylinder radii different from
  Fig.~\ref{fig:intro}. The shaded regions highlight the {\it
    frequency} gap (with complex {\it momentum} solutions inside) in
  the left panel, and the {\it momentum} gap (with complex {\it
    frequency} solutions inside) in the right panel. Note the space -
  time duality of the two regions.}
\label{fig:comparison}
\end{figure}

In the 2D problem, the critical value of the cylinder radius $R_c$,
above which folded bands are formed, may depend on the direction in
momentum space. In fact, Fig.~\ref{fig:2d} shows that for instance for
$R=0.27a$, the two lowest bands are already folded in the $\Gamma M$
direction.

It turns out that one can find a simple criterion for the appearance
of folded bands. In fact, from the Maxwell equations in the frequency
domain, $\textrm{rot} {\vec E}=i\omega {\vec B}$ and $\textrm{rot}{\vec
  H}{^\ast}=i\omega^\ast {\vec D}^\ast$, after multiplication by ${\vec
  H}^\ast$ and ${\vec E}$, respectively, follows the identity
\begin{equation}
i\omega_1 u_{-} + \omega_2 u_{+} + \textrm{div}{\vec S}=0,
\label{eq:identity}
\end{equation}
where $u_\pm={\vec E}\cdot{\vec D}^\ast \pm {\vec B}\cdot{\vec
  H}^\ast$ and ${\vec S}={\vec E}\times{\vec H}^\ast$. Note that this
identity holds also in dispersive and/or lossy media and it represents
a generalization of Poynting's theorem for harmonic fields
\cite{Jackson}. In our case of lossless media the quantities $u_\pm$
are both real. Let us integrate Eq.~(\ref{eq:identity}) over the area
of an elementary cell of the photonic crystal. The Bloch boundary
conditions Eq.~(\ref{eq:Bloch}) imply that $\int_{\rm cell} d^2x
\textrm{div}{\vec S}=0$. Therefore $i\omega_1 U_{-} + \omega_2
U_{+}=0$ must hold, where $U_{\pm}=\int_{\rm cell} d^2x
\left[\varepsilon |{\vec E}|^2 \pm \mu |{\vec H}|^2\right]$.
Considering the real part of this condition for ${\cal P}{\cal T}$
symmetry-breaking solutions with $\omega_2\neq 0$, we find that
$U_{+}=0$ must hold, as confirmed for the 1D problem in
Fig.~\ref{fig:field}(b). Similar criteria for the occurence of folded
bands have been found in \cite{chen} by different reasoning.  Note
that at the verge of ${\cal P}{\cal T}$ symmetry breaking, the group
velocity diverges and $U_{+}$ vanishes. Explicit calculation for the
1D problem shows that the real part of ${\vec S}$ stays finite here,
see Fig.~\ref{fig:field}(c).

Before concluding let us observe that folded bands are closely related
to the frequency gap in periodic systems. In fact, it is well known
that within the frequency gap, there exist only solutions with a
complex wave-vector. On the other hand, folded bands imply the
existence of a {\it momentum gap}, inside which there appear only
states with a complex frequency. Thus the momentum gap is a space -
time dual of the frequency gap, see Fig.~\ref{fig:comparison}.

In conclusion, we have explained the physical origin of folded bands,
or momentum gaps, in spectra of metamaterial crystals. Two ingredients
are necessary for their appearance: first, the boundary-value problem
has to be essentially non-Hermitian, and second, the possible reality
of its spectrum has to be guaranteed by additional symmetry, such as
the ${\cal P}{\cal T}$ symmetry in the metamaterial case. In systems
fulfilling these assumptions, momentum gaps should be commonplace.

\medskip
We thank V.~Balek and M.~Moj\v{z}i\v{s} for helpful discussions.  This
work was supported by the Slovak Research and Development Agency under
the contract No. APVV-0108-11 and by the Agency VEGA under the
contracts No.~1/0372/13 and No.~1/0904/15.


\begin{thebibliography}{99}

\bibitem{sakoda} K.~Sakoda, Optical Properties of Photonic Crystals,
  Berlin, Heidelberg: Springer (2005).

\bibitem{chen} P.~Y.~Chen \textsl{et al}, {New J. Phys.} \textbf{13}
  053007 (2011).

\bibitem{pm} P.~Marko\v{s}, unpublished. The method has been described
  in P.~Marko\v{s}, preprint arXiv:1501.05125.

\bibitem{john} It is worth pointing out that folded bands have been
  observed also earlier, see, e.g., D.~Toader and S.~John,
  Phys. Rev. E \textbf{70} 046605 (2004) and D.~Hermann \textsl{et
    al}, {Phys. Rev. B} \textbf{77} 035112 (2008). However, in those
  papers systems with frequency-dependent material parameters were
  studied. Therefore, from the mathematical point of view, those
  authors did not study eigenvalue problems, and the presence of
  folded bands in spectra did not lead to the sort of questions which
  we address.

\bibitem{joan-pc} J.~D.~Joannopoulos {\it et al.}, 
Photonic Crystals: Molding the Flow of Light, 2nd ed. Princeton:
Princeton University Press (2008).

\bibitem{note}Since we are dealing with a metamaterial crystal,
  Hermitization of the wave equation for $\vec{E}$ in terms of the
  substitution $\vec{E}=\vec{e}/\sqrt{\varepsilon}$ is plagued by the
  same problem.

\bibitem{Nefedov02}I.~S.~Nefedov and S.~A.~Tretyakov, Phys. Rev. E
  {\bf 66}, 036611 (2002).

\bibitem{Li03}Jensen Li {\it et al.}, 
Phys. Rev. Lett. {\bf 90}, 083901 (2003).

\bibitem{Wu03}Liang Wu, Sailing He, and Linfang Shen, Phys. Rev. B
  {\bf 67}, 235103 (2003)

\bibitem{Bria04}D. Bria {\it et al.}, Phys. Rev. E {\bf 69}, 066613
  (2004).

\bibitem{Bender05}For a review, see C.~M.~Bender, Contemporary Physics
  {\bf 46}, 277 (2005).

\bibitem{Yeh05}P. Yeh, Optical Waves in Layered Materials, Hoboken,
  New Jersey: Wiley (2005).

\bibitem{proof}To this end it suffices to note that for frequencies
  $\omega_n=n\pi/\tau_{+}$, where $n$ is an integer, we have
  $|f(\omega_n)|\geq 1$. Moreover, $f(\omega_n)$ exhibits even-odd
  oscillations with $n$. But since the second term in $f(\omega)$
  oscillates with a longer period, in between $\omega_n$ and
  $\omega_{n+1}$ there will be at most one extreme of $f(\omega)$, and
  therefore $|f(\omega^\ast)|\geq|f(\omega_n)|\geq 1$.

\bibitem{zeron}Also the much debated zero-$\bar{n}$ gaps \cite{Li03}
  are associated with a special case of folded bands.  In fact, the
  zero-$\bar{n}$ condition is equivalent to $\tau_{+}=0$ and in this
  case Eq.~(\ref{eq:1d}) has solutions only for $q=0$ and
  $\omega\tau_{-}=2m\pi$, where $m$ is an integer.

\bibitem{optics}In previous work, the concept of ${\cal P}{\cal T}$
  symmetry has been exploited in optics from a quite different
  perspective, making use of the formal analogy between quantum
  mechanics and the paraxial approximation. The ${\cal P}{\cal T}$
  symmetry was achieved by judicious spatial modulation of loss and
  gain, see e.g. C.~E.~R\"{u}ter {\it et al.}, Nat. Phys. {\bf 6}, 192
  (2010). The closest in spirit to ours is the paper L.~Ge and
  H.~E.~T\"{u}reci, Phys. Rev. A {\bf 88}, 053810 (2013), which also
  studies metamaterial crystals (with different symmetry) using the
  concept of ${\cal P}{\cal T}$ symmetry. That paper is limited only
  to 1D systems, however, and, more importantly, it does not study the
  eigenvalue problem.

\bibitem{Krejcirik06}D. Krej\v{c}i\v{r}\'{i}k, H. B\'{i}la, and
  M. Znojil, J. Phys. A: Math. Gen. {\bf 39}, 10143 (2006).

\bibitem{Mostafazadeh11}For a recent example, see A.~Mostafazadeh,
  J. Phys. A: Math. Theor. {\bf 44}, 375302 (2011).

\bibitem{note2}In the ${\cal P}{\cal T}$ symmetry-breaking sector, we
  plot $|H(x)|$ for only one solution for every value of $q$. The
  minima of $|H(x)|$ shift to the left from the center of layer $a$.
  The other solution (not shown), which corresponds to a
  complex-conjugate frequency, exhibits an opposite shift.

\bibitem{Jackson}J.~D.~Jackson, Classical Electrodynamics, 3$^{\rm
  rd}$ ed., New York: John Wiley (1999). 

\end{thebibliography}
\end{document}